\documentclass[pdflatex,sn-mathphys-num]{sn-jnl}

\usepackage{graphicx}
\usepackage{multirow}
\usepackage{amsmath,amssymb,amsfonts}
\usepackage{amsthm}
\usepackage{mathrsfs}
\usepackage[title]{appendix}
\usepackage{xcolor}
\usepackage{textcomp}
\usepackage{manyfoot}
\usepackage{booktabs}
\usepackage{algorithm}
\usepackage{algorithmicx}
\usepackage{algpseudocode}
\usepackage{listings}
\usepackage{siunitx}
\usepackage{orcidlink}

\theoremstyle{thmstyleone}

\theoremstyle{thmstyletwo}

\theoremstyle{thmstylethree}

\begin{document}

\title[Article Title]{AOM-based ultra-low noise laser intensity control up to the MHz range}

\author*[1]{Christian Partes${}^{\orcidlink{0009-0001-9466-8495}}$}

\author[1]{Jonas Auch${}^{\orcidlink{0009-0005-5863-8239}}$}

\author[1]{Eduard Heidt${}^{\orcidlink{0009-0007-1294-5043}}$}

\author[1]{Kirill Karpov${}^{\orcidlink{0009-0000-2701-5593}}$}

\author[1]{Florian Kiesel${}^{\orcidlink{0009-0009-0515-3069}}$}

\author[2]{Christian~Gro\ss${}^{\orcidlink{0000-0003-2292-5234}}$}\email{christian.gross@uni-tuebingen.de}

\affil[1]{Physikalisches Institut, Eberhard Karls Universit\"at T\"ubingen, 72076 T\"ubingen, Germany}

\affil[2]{Physikalisches Institut and Center for Integrated Quantum Science and Technology, Eberhard Karls Universit\"at T\"ubingen, 72076 T\"ubingen, Germany}

\abstract{High-power laser sources that exhibit low relative intensity noise and allow simultaneous dynamic control of their light level are required for a broad range of applications in various fields of physics.
	Acousto-optic modulators (AOMs) are widely used for active power stabilization and regulation due to their simple drive electronics requirements and high optical power handling capability in free-space.
	However, the rather slow propagation speed of the sound wave within the AOM crystal typically limits their control bandwidth to a few hundred kHz.
	In this work, we present a novel AOM-based control system that is capable of significantly suppressing intensity noise of high-power lasers up to the MHz range.
	By combining two standard feedback loops with one feedforward control branch and optimizing the beam path in the AOM crystal, ultra-low relative intensity noise levels down to \qty{-155}{\dB\per\Hz} even at several hundred kHz are achieved.
	Our results are relevant for applications that require ultra-low intensity noise at Fourier frequencies up to the MHz range, such as optical lattice experiments with light ultracold atoms.}

\keywords{Intensity control, AOM, feedforward, feedback, relative intensity noise}

\maketitle

\section{Introduction}
Stabilizing and controlling the power of laser sources is required for a wide variety of physical applications \cite{tricot_power_2018, wang_reduction_2020}.
Lasers with low intensity fluctuations are needed to enhance the performance of atomic magnetometers \cite{duan_light_2015}, to achieve narrow linewidths of random Raman fiber lasers \cite{dong_high_2018}, or to increase the precision of atomic clocks \cite{dawel_high-stability_2026}.
Other applications that are sensitive to optical power instabilities include optical communication \cite{villenas_optical_2026} and gravitational wave detection \cite{takahashi_stabilization_2008, barr_laser_2005}.
In the context of quantum information and simulation, intensity noise degrades the fidelity of laser-driven quantum gates \cite{jiang_sensitivity_2023, day_limits_2022} and induces parametric heating in optical dipole traps \cite{gehm_dynamics_1998, sun_influence_2020, wang_reduction_2020}.
The latter is of paramount importance for optical lattice experiments with light ultracold atoms as the involved high trapping frequencies require the relative intensity noise (RIN) spectrum of the lattice laser source to remain low up to the MHz range \cite{blatt_low-noise_2015, mazurenko_implementation_2019}.
For the generation of such optical lattices, commercial lasers exhibiting low RIN approaching \qty{-150}{\dB\per\Hz} at Fourier frequencies above \qty{100}{\kilo\Hz} are available at \qty{1064}{\nano\meter} \cite{mazurenko_implementation_2019}.
However, this wavelength is not always usable due to conflicting requirements, such as the need to use (anti-)magic \cite{anisimovas_semisynthetic_2016, holman_mid-infrared_2026} or tune-out wavelengths \cite{de_martino_dissipationless_2025}.

In the past, a variety of different approaches have been developed to reduce laser intensity noise.
While some methods rely on passive stabilization by shielding the laser from temperature and pressure fluctuations as well as mechanical vibrations \cite{talvitie_passive_1997}, most approaches involve active stabilization with feedback loops.
Although different techniques are available to tune the laser intensity \cite{ottaway_frequency_2000, barr_laser_2005, karami_novel_2025}, this is typically done using AOMs \cite{tricot_power_2018, phrompao_real-time_2019, blatt_low-noise_2015, kim_laser_2007} or electro-optic modulators (EOMs) \cite{michael_broadband_2015, robertson_intensity_1986, liu_long-term_2013, nelson_relative_2008, wang_reduction_2020, wang_relative_2024, kwee_new_2011} as actuators of the control system.
With an EOM, an exceptionally high stabilization bandwidth of up to \qty{10}{\mega\Hz} has already been reached \cite{nelson_relative_2008}.
Even higher bandwidths are achievable by implementing feedforward intensity stabilization systems on EOM-based platforms.
In atomic, molecular, and optical (AMO) physics experiments, feedforward is commonly used for phase and frequency control \cite{chao_robust_2025, li_active_2022}, and has been demonstrated to stabilize the power of consecutive laser pulses \cite{wang_relative_2024} as well as the intensity of a \qty{40}{\milli\watt} continuous-wave laser in the GHz regime \cite{michael_broadband_2015}.

Although EOMs exhibit much faster response times than AOMs, the latter are commonly used in laboratory laser setups due to their simple driving requirements and high power handling capability.
However, their stabilization bandwidth is usually limited to several hundred kHz \cite{tricot_power_2018, xu_quantum_2024, wang_reduction_2020}, preventing their usage for applications that are sensitive to high-frequency intensity noise.

In this work, we present the implementation of a fast intensity control platform based on two AOMs, which is capable of both adjusting the average power level of our laser source over the full dynamic range and reducing the RIN spectrum even beyond \qty{1}{\mega\Hz}.
Our method is suitable for optical powers up to several watts, with the exact power limit determined by the wavelength-dependent damage thresholds of the AOMs and the optical fiber employed.
Additionally, the average light level can be ramped from zero to its maximum level within several tens of milliseconds.
Non-linearities in the AOM response are accounted for by a gain-scheduling algorithm implemented on a Field-Programmable Gate Array (FPGA).
Intensity noise reduction by more than \qty{30}{\dB} down to \qty{-155}{\dB\per\Hz} even at several hundred kHz is achieved by a novel combination of feedback and feedforward in a single AOM.
In the following, we demonstrate the RIN suppression performance of our control platform and verify its ability to ramp the average laser intensity on the time scale specified above.

\section{Experimental setup}
\label{ch: Experimental setup}

\begin{figure}
	\centering \includegraphics{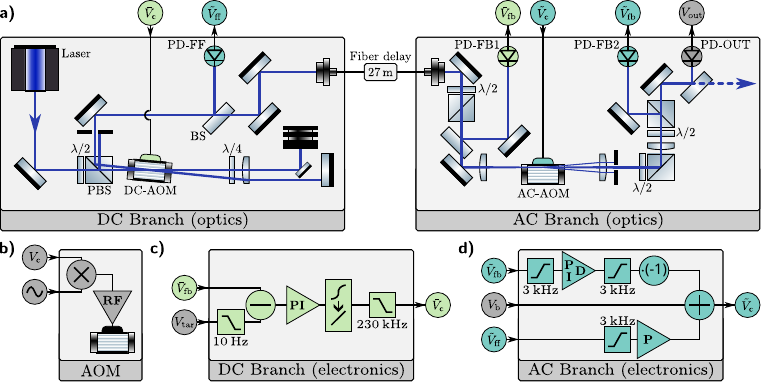}
	\caption{
		Overview of the experimental setup.
		Dark blue lines represent the laser beam path, black lines indicate electronic connections.
		Components and signals that are part of the DC (AC) branch are marked in light (mint) green. a)~Schematic representation of the beam path and the positioning of the photodiodes.
		The feedback signal and the control signal of the DC branch are labeled with $\bar{V}_\mathrm{fb}$ and $\bar{V}_\mathrm{c}$, respectively.
		In the AC branch, the feedforward signal~$\tilde{V}_\mathrm{ff}$ and the fast feedback signal~$\tilde{V}_\mathrm{fb}$ are processed to obtain the AC control voltage~$\tilde{V}_\mathrm{c}$.
		The out-of-loop signal~$V_\mathrm{out}$ serves as an independent measurement of the RIN.
		It is obtained from the output of the control system indicated by the dashed arrow.
		For simplicity, some components that are not necessary for the understanding of the setup are left out. b)~Electronics driving the AOMs.
		An \qty{80}{\mega\Hz} driving signal is modulated by the applied control voltage~$V_\mathrm{c}$ using an RF mixer.
		The resulting output is amplified and sent to the respective AOM. c)~Signal path of the slow DC branch.
		The low-pass filtered target voltage~$V_\mathrm{tar}$ corresponds to the setpoint of the feedback signal~$\bar{V}_\mathrm{fb}$.
		The difference between these two signals is processed by a digital PI-controller.
		Its output is modified to compensate for non-linearities in the DC-AOM response and is subsequently low-pass filtered, resulting in the control voltage~$\bar{V}_\mathrm{c}$. d)~Signal path of the fast AC branch.
		Both input voltages~$\tilde{V}_\mathrm{fb}$ and $\tilde{V}_\mathrm{ff}$ are high-pass filtered.
		While the feedback signal~$\tilde{V}_\mathrm{fb}$ is processed by a PID controller and is subsequently inverted, the amplitude of the feedforward signal~$\tilde{V}_\mathrm{ff}$ is adjusted using a variable gain amplifier.
		The two output voltages are added to a bias offset voltage~$V_\mathrm{b}$ to obtain the control output~$\tilde{V}_\mathrm{c}$. }
	\label{fig: Experimental setup}
\end{figure}

The laser source whose beam intensity is to be stabilized is a titanium-sapphire laser tuned to \qty{841}{\nano\meter}.
The intensity control system is divided into a slow (DC) and a fast (AC) branch \cite{liu_long-term_2013, xu_quantum_2024,wang_reduction_2020}.
While the DC branch regulates the average intensity to an external setpoint and suppresses low-frequency power fluctuations, the AC branch is provided with a high-pass filtered intensity signal and reduces the RIN spectrum at Fourier frequencies higher than \qty{3}{\kilo\Hz}.

The optical setup of the intensity control system is shown in \autoref{fig: Experimental setup}~a).
Around \qty{170}{\milli\watt} of the light emitted by the laser is sent through a polarizing beamsplitter (PBS) followed by a double-pass AOM configuration \cite{donley_double-pass_2005}.
The AOM (DC-AOM: Gooch~\&~Housego AOMO 3080-122) used in this setup serves as the actuator of the DC branch.
Due to the back reflection of the diffracted laser beam, the pointing direction of the outgoing beam does not depend on the DC-AOM driving frequency, as would be the case for a single-pass setup.
The maximum diffraction efficiency reached by the double-pass setup is around $75\,\%$.
Having passed the DC-AOM configuration, the laser light is coupled into a \qty{27}{\meter} single-mode fiber delay line.
For high-power applications, this delay line can be replaced with a photonic crystal fiber.
After the outcoupler, the light's polarization is projected onto the vertical polarization axis using a PBS.
The preceding half-wave plate (HWP) is adjusted to maximize the transmission through the PBS.
Subsequently, the light passes through a second AOM (AC-AOM: Gooch~\&~Housego AOMO 3080-122).
In contrast to the DC-AOM, its $0^\mathrm{th}$~diffraction order is retained rather than the first, resulting in a significantly higher transmission efficiency compared to the double-pass setup.
The reaction time of the AC-AOM can be optimized by minimizing the propagation time required by the sound wave within its crystal to reach the laser beam and to cover its entire diameter \cite{xu_quantum_2024}.
To this end, the laser is focused through the AOM using a \qty{150}{\milli\meter}-lens.
A horizontal beam waist of around \qty{50}{\micro\meter} in the focal plane has been experimentally found to maximize the bandwidth of the AOM.
For our specific wavelength, the corresponding average power density in the focal plane does not exceed the specified AOM damage threshold for optical powers up to \qty{2}{\watt}.
Moreover, the beam path was carefully adjusted to minimize its distance to the AOM's acoustic transducer while clipping of the laser beam at the crystal edge was avoided.
In this configuration, reaction times down to \qty{65}{\nano\s} are achieved (see methods \ref{ch: AOM response}), while more than 92\,\% of the optical power is transmitted by the undriven AOM.
The corresponding losses can be greatly reduced by moving the beam path further away from the acoustic transducer at the expense of slightly higher reaction times.
After being focused through the AC-AOM, the laser beam is collimated again and sent through another PBS compensating for possible polarization noise and drifts.
Finally, the outgoing beam is again focused by a \qty{150}{\mm}-lens and split in a variable ratio by an HWP-PBS combination.
Two photodiodes (PDs) are placed in the focal planes of both the reflective and the transmissive output of the PBS.
While the PD in the reflected beam (PD-FB2) provides a feedback signal~$\tilde{V}_\mathrm{fb}$, the diode in the transmitted beam (PD-OUT) is placed in the usable output of the control system and generates a voltage~$V_\mathrm{out}$ serving as an out-of-loop measurement of the RIN spectrum and the intensity level \cite{wang_reduction_2020, liu_long-term_2013, kim_laser_2007}.
Using a 16-bit oscilloscope, the RIN is measured up to \qty{5}{\mega\Hz} with a measurement bandwidth of \qty{305}{\Hz}.
Both PDs are placed at a significant angle with respect to the propagation axis of the beam to avoid the influence of reflections on the RIN measurements.
Additionally, signal distortions caused by interference effects are prevented by removing the glass covers of their photosensitive areas \cite{mazurenko_implementation_2019}.
The transimpedance amplification circuit of the PDs is custom-designed to allow the detection of RIN levels below \qty{-150}{\dB\per\Hz} (see methods \ref{ch: Low-noise Photodiodes}).
Due to its low transimpedance gain of \qty{1.5}{\kilo\ohm}, both PDs are exposed to an optical power of around \qty{13.5}{\milli\watt} to achieve output voltages of \qtyrange[range-units=single,range-phrase=\,-\,]{7.5}{8.2}{\volt}.

In order to drive the AOMs, RF mixers are used to modulate \qty{80}{\mega\Hz} RF driving signals generated by voltage-controlled oscillators (VCOs) with the control signals $V_\mathrm{c}$ provided by the two control branches, see~\autoref{fig: Experimental setup}~b).
The outputs of the RF mixers are amplified and sent to the corresponding AOMs \cite{mazurenko_implementation_2019}.
In this configuration, the fraction of the optical power diffracted by the modulators depends on their applied control signal~$V_\mathrm{c}$:

\begin{equation}
	\label{eq: AOM intensity relation}
	\frac{P_\mathrm{out}}{P_\mathrm{in}}\propto\sin^2\left(\sqrt{\frac{P_\mathrm{RF}(V_\mathrm{c})}{P_\mathrm{sat}}}\right),
\end{equation}
where $P_\mathrm{in}$ and $P_\mathrm{out}$ are the incoming and diffracted beam powers, $P_\mathrm{RF}$ is the power of the applied RF wave, and $P_\mathrm{sat}$ is the saturation power of the AOMs \cite{xu_quantum_2024}.
For the AC-AOM, the cables employed in this setup are chosen to be as short as possible to reduce the time delay introduced by signal propagation.
Equation~\eqref{eq: AOM intensity relation} implies that the intensity transmitted by the DC control system increases with the applied control voltage~$\bar{V}_\mathrm{c}$ as only the first diffraction order of the DC-AOM is retained.
In contrast, the intensity transmitted by the AC control branch decreases with the control signal~$\tilde{V}_\mathrm{c}$ since the $0^\mathrm{th}$~diffraction order of the AC-AOM is used.

\section{Low-frequency intensity control}
\label{ch: Low-frequency intensity control}

The DC intensity control system is implemented on the FPGA-based Red Pitaya (RP) platform using a modified version of the electronic setup presented in \cite{preuschoff_digital_2020}.
A schematic overview of the DC control branch is shown in \autoref{fig: Experimental setup}~c).
Its feedback signal is the voltage~$\bar{V}_\mathrm{fb}$ generated by a PD (PD-FB1, transimpedance gain: \qty{20}{\kilo\ohm}) located in front of the AC-AOM.
It is exposed to an optical power of around \qty{250}{\micro\watt}, which is split from the main beam using a beam sampler (BS, see~\autoref{fig: Experimental setup}~a)).
The target setpoint for the intensity is given by an external voltage~$V_\mathrm{tar}$.
The latter is low-pass filtered using a \qty{10}{\Hz} RC-filter to avoid rapid intensity variations, which would impair the stability of the AC control branch.
However, higher DC modulation bandwidths are achievable at the expense of reduced RIN suppression performance in the low-kHz regime (see~\autoref{ch: DC level modulation}).
The difference between the signals~$\bar{V}_\mathrm{fb}$ and~$V_\mathrm{tar}$ is processed by a PI-controller implemented on the FPGA, allowing the proportional and integral gains~$G_\mathrm{P}$ and $G_\mathrm{I}$ to be adjusted remotely.
For this work, if not specified differently, $G_\mathrm{P}=0$ and $G_\mathrm{I}=3000$ (internal units) are chosen, effectively disabling the proportional controller.
In this configuration, the DC branch provides sufficient control of the light intensity to track the target voltage~$V_\mathrm{tar}$, while the formation of a significant servo bump is avoided.
Finally, the resulting control voltage~$\bar{V}_\mathrm{c}$ is low-pass filtered with a \qty{230}{\kilo\Hz} RC-filter to suppress the negative effect of high-frequency noise spikes introduced by the DC branch electronics.

\begin{figure}
	\centering \includegraphics{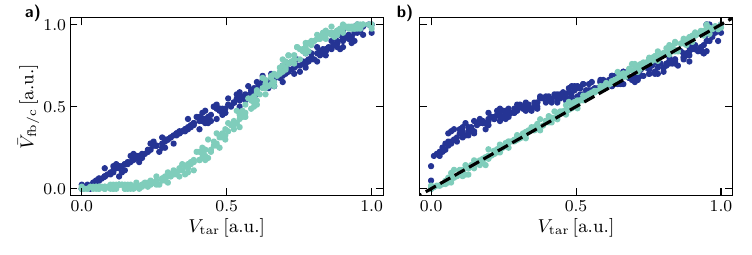}
	\caption{a)/b) Dependence of the control voltage~$\bar{V}_\mathrm{c}$ (dark blue) and the PD-FB1 output voltage~$\bar{V}_\mathrm{fb}$ (mint green) on the target voltage~$V_\mathrm{tar}$ without~(a)~and with~(b)~active linearization. The signals have been offset and normalized to vary between 0 and 1 for better reading. The black dashed line represents an ideal linear dependence as a guide for the eye.} 
	\label{fig: Linearization}
\end{figure}

Generally, AOM-based intensity control systems suffer from a non-linear dependence of the outgoing beam intensity on the control voltage~$V_\mathrm{c}$, making the effective proportional and integral gains $G_\mathrm{P}$ and $G_\mathrm{I}$ depend on the target voltage~$V_\mathrm{tar}$.
This non-linearity is caused by both the diffraction efficiency in equation \eqref{eq: AOM intensity relation} and the non-linear response of the RF-mixer, resulting in reduced control performance and possible instabilities \cite{mazurenko_implementation_2019}.
The dependence of the total diffracted beam power on the applied control voltage~$\bar{V}_\mathrm{c}$ can be measured by disconnecting the feedback signal~$\bar{V}_\mathrm{fb}$ from the DC controller and setting $G_\mathrm{P}=1$ and $G_\mathrm{I}=0$.
Using these settings, the control signal calculated by the digital PI-controller equals the target voltage~$V_\mathrm{tar}$ except for minor electronic offsets on the order of a few mV.
The setpoint~$V_\mathrm{tar}$ is swept from \qty{-10}{\milli\volt} to \qty{380}{\milli\volt}, and the applied control voltage~$\bar{V}_\mathrm{c}$ and the PD-FB1 signal~$\bar{V}_\mathrm{fb}$ are recorded using an oscilloscope.
The results are shown in \autoref{fig: Linearization}~a).
As expected, $\bar{V}_\mathrm{c}$ follows $V_\mathrm{tar}$ linearly, while the slope of the PD-FB1 signal~$\bar{V}_\mathrm{fb}$ varies significantly throughout the measured interval.

To counteract this effect, a linearization algorithm based on the gain-scheduling approach is implemented on the FPGA.
The actual control output~$\bar{V}_\mathrm{c}$ is obtained by applying the inverse~$f^{-1}$ of the non-linear function~$f(\bar{V}_\mathrm{c}) = \bar{V}_\mathrm{fb}(\bar{V}_\mathrm{c})$ to the control signal calculated by the standard PI-controller.
On the FPGA,~$f^{-1}$ is implemented by a piecewise linear approximation (PLA) with 9 independent partitions.
The measurement of the function~$f(\bar{V}_\mathrm{c})$ and the calculation of the slope parameters required for the PLA are performed using an automatic scanning algorithm.
With linearization enabled, the measurement of the PD-FB1 output voltage $\bar{V}_\mathrm{fb}$ and the control voltage~$\bar{V}_\mathrm{c}$ described above was repeated.
In \autoref{fig: Linearization}~b), the results are presented.
The control voltage~$\bar{V}_\mathrm{c}(V_\mathrm{tar})$ follows the inverse of the PD-FB1 output voltage~$\bar{V}_\mathrm{fb}(V_\mathrm{tar})$ in the non-linearized case (see \autoref{fig: Linearization}~a)).
Consequently, with linearization enabled, the slope of the latter remains approximately constant throughout the entire scan.
The root mean square deviation (RMSD) between the normalized PD-FB1 signal~$\bar{V}_\mathrm{fb}(V_\mathrm{tar})$ and the ideal linear behavior indicated by the black dashed line is 0.024 (0.138) for the linearized (non-linearized) case, corresponding to an improvement by a factor of 5.75.
The effect of the algorithm could be further improved by increasing the number of partitions of the PLA at the expense of an additional demand for hardware resources on the FPGA.

Since the DC branch can operate as a stand-alone intensity control system, its optimized RIN suppression performance is determined.
For this purpose, both the RIN spectra of the free-running laser and the intensity-stabilized laser regulated by the DC control branch are recorded.
Note that they are obtained from the high-pass filtered AC~output~$V_\mathrm{out}$ of PD-OUT (cutoff frequency: \qty{3}{\kilo\Hz}, see methods \ref{ch: Low-noise Photodiodes}).
For the measurement, the constant target voltage~$V_\mathrm{tar}$ was adjusted to its optimal value~$V_\mathrm{tar}^\mathrm{opt} = 950\,\mathrm{mV}$ to maximize the DC branch transmission efficiency while the controller remained capable of counteracting low-frequency polarization drifts.
To optimize control performance, the gains were set to $G_\mathrm{P} = 0.02$ and $G_\mathrm{I} = 60000$, and the \qty{230}{\kilo\Hz} low-pass filter at the control output~$\bar{V}_\mathrm{c}$ was removed.
For the measurement of the free-running laser's noise spectrum, a small integrator-term ($G_\mathrm{I}=10$) was kept in order to avoid DC intensity drifts.
The results are presented in \autoref{fig: Final Rin}.
The RIN level of the controlled laser reaches \qty{-129}{\dB\per\Hz} at \qty{13}{\kilo\Hz}, while the noise reduction bandwidth is around \qty{90}{\kilo\Hz}.
The latter is limited by the AOM response time and the signal processing time on the FPGA, both of which amount to several hundred ns.
This implies the necessity of the additional AC control branch for high-frequency intensity stabilization up to the MHz-range.

\section{High-frequency intensity control}
\label{ch: AC branch}
The main contributor to noise suppression at Fourier frequencies above \qty{10}{\kilo\Hz} is the fast AC controller depicted in \autoref{fig: Experimental setup}~d).
This controller consists of a combination of a standard feedback system and a feedforward branch.
For the implementation of the feedback stage, the intensity of the laser beam behind the AC-AOM is measured, and the corresponding output voltage of PD-FB2 is high-pass filtered (see methods \ref{ch: Low-noise Photodiodes}).
The resulting feedback signal~$\tilde{V}_\mathrm{fb}$ is sent to the non-inverting input of the fast branch of the Toptica FALC110 PID controller (FALC), providing the feedback control voltage~$\tilde{V}_\mathrm{FALC}$.
The inverting input of the FALC is left floating.
In order to have full control over the DC setpoint of the AC-AOM control voltage~$\tilde{V}_\mathrm{c}$, the FALC control output~$\tilde{V}_\mathrm{FALC}$ is again high-pass filtered with a \qty{3}{\kilo\Hz} RC-filter and added to an external bias voltage~$V_\mathrm{b}=160\,\mathrm{mV}$ using a custom analog summing amplifier (see methods \ref{ch: Summing amplifier}).
Additionally, inverting the control signal~$\tilde{V}_\mathrm{FALC}$ is necessary as the intensity transmitted by the AC control system decreases with increasing control voltage~$\tilde{V}_\mathrm{c}$.
Due to the applied bias voltage~$V_\mathrm{b}$, the power after the AC-AOM is reduced by around 3\,\% compared to the undriven AOM.
For the implementation of the feedforward control system, the light intensity directly in front of the fiber delay line is measured using a commercial PD (PD-FF: Thorlabs PDA10A2).
The generated voltage~$\tilde{V}_\mathrm{ff}$ (typically \qtyrange[range-units=single,range-phrase=\,-\,]{9}{10}{\volt}) is sent through a \qty{7}{\meter} SMA cable.
This cable length was chosen as it corresponds to the shortest possible length that could be implemented in our physical setup and resulted in the best control performance.
The feedforward signal~$\tilde{V}_\mathrm{ff}$ is high-pass filtered with a \qty{3}{\kilo\Hz} RC-filter and sent to a variable gain amplifier to adjust its signal amplitude to match the response efficiency of the AC-AOM.
The resulting control signal is added to the inverted feedback signal~$\tilde{V}_\mathrm{FALC}$ and the bias voltage~$V_\mathrm{b}$, resulting in the final control voltage~$\tilde{V}_\mathrm{c}$.
This procedure combines two kinds of control mechanisms on one single actuator, reducing the complexity of the optical setup and enlarging the total transmission efficiency in comparison to two separate controllers.
Additionally, environmental drifts, which are reported to reduce the performance of feedforward systems \cite{chao_robust_2025}, lose significance due to the presence of more robust feedback control.

\begin{figure}
	\centering \includegraphics{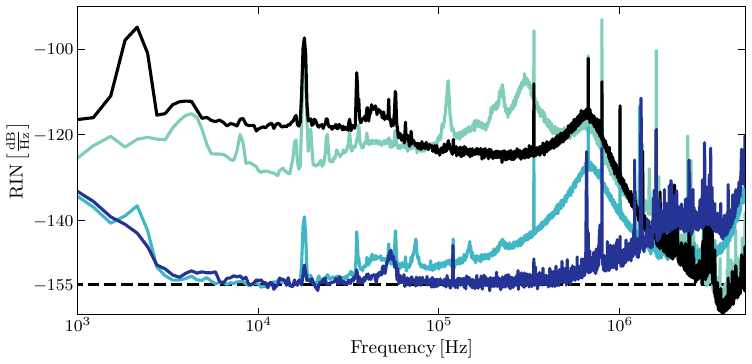}
	\caption{
		RIN spectrum of the free-running laser (black) and the laser stabilized by the complete control system (dark blue).
		Additionally, the spectra obtained by using only the slow DC branch (mint green) and the fast AC branch without feedforward (light blue) are shown.
		The horizontal, dashed line at \qty{-155}{\dB\per\Hz} indicates the lowest achieved RIN level.}

	\label{fig: Final Rin}
\end{figure}

The RIN suppression performance of the combined DC and AC control system with and without active feedforward stabilization is shown in \autoref{fig: Final Rin}.
As before, the target setpoint~$V_\mathrm{tar}$ was set to~$V_\mathrm{tar}^\mathrm{opt}$ for all measurements.
With feedforward enabled, the RIN spectrum remains below \qty{-145}{\dB\per\Hz} up to \qty{1.2}{\mega\Hz} except for several noise spikes, which might be caused by spurious electronic noise.
The closed-loop control bandwidth is approximately \qty{1.5}{\mega\Hz}, with the servo bump located at around \qty{4.8}{\mega\Hz}.
To the best of our knowledge, this corresponds to one of the highest control bandwidths ever achieved on an AOM-based platform.
The spectrum reaches \qty{-155}{\dB\per\Hz} from \qty{10}{\kilo\Hz} to \qty{200}{\kilo\Hz}, which is in agreement with our estimate of the achievable RIN limit~$S_\mathrm{min}=\,$\qty{-158}{\dB\per\Hz} imposed by the used PDs (see methods \ref{ch: Low-noise Photodiodes}).
The difference of \qty{3}{\dB} between the calculated and the actual noise floor might be explained by electronic noise induced by the summing amplifier or other electronic components involved.
Total noise reduction stays at or above \qty{29}{\dB} up to \qty{700}{\kilo\Hz}.
At this specific frequency, \qty{23}{\dB} of noise suppression is achieved by the feedforward branch, which thus constitutes the most important part of the control system at high frequencies.
By fine-tuning the length of the used SMA cable to match the electronic and optical delays involved in the feedforward system, noise reduction at high frequencies might be further improved.
Additionally, the flattening of the RIN spectrum in the kHz-regime implies the feasibility of even lower RIN levels by using PDs with lower electronic noise levels and lower shot noise limits.

\section{DC level modulation}
\label{ch: DC level modulation}
\begin{figure}
	\centering \includegraphics{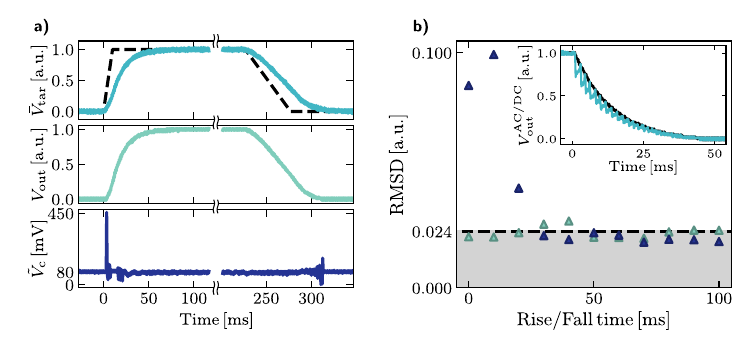}
	\caption{
		a) Time dependence of the PD-OUT output voltage~$V_\mathrm{out}$ (mint green) and the AC  control voltage~$\tilde{V}_\mathrm{c}$ (dark blue) during an increase and decrease of the target signal $V_\mathrm{tar}$.
		The low-pass filtered setpoint voltage $\bar{V}_\mathrm{tar}$ is shown in light blue.
		For reference, the black dashed line indicates an unfiltered pulse signal with a rise (fall) time of \qty{10}{\milli\s} (\qty{50}{\milli\s}).
		The output voltage~$V_\mathrm{out}$ and the filtered target voltage~$\bar{V}_\mathrm{tar}$ have been normalized and offset to vary between 0 and 1.
		At time~$t=0\,\mathrm{s}$, the ramp of the target voltage~$V_\mathrm{tar}$ begins. b) Root mean square deviation (RMSD) between the intensity curves during signal ramps with~($V_\mathrm{out}^{\mathrm{AC}}(t)$) and without~($V_\mathrm{out}^{\mathrm{DC}}(t)$) active AC control.
		The data points for the rising (falling) edges are shown in mint green (dark blue).
		The black dashed line indicates an RMSD of 0.024, corresponding to the average RMSD of the measured rising intensity curves.
		The shaded area below this line marks the related noise-induced lower measurement limit.
		The inset shows the waveforms~$V_\mathrm{out}^{\mathrm{AC}}(t)$ (light blue) and~$V_\mathrm{out}^{\mathrm{DC}}(t)$ (black) for an applied fall time of~\qty{0}{\milli\s}.
		The shown signals have been normalized and offset to vary between 0 and 1.}
	\label{fig: DC Mod}
\end{figure}
The behavior of the laser intensity during a moderate ramp of the target level~$V_\mathrm{tar}$ is demonstrated by applying pulses with a maximum level of~$V_\mathrm{tar}=V_\mathrm{tar}^\mathrm{opt}$, a rise time of \qty{10}{\milli\s}, and a fall time of \qty{50}{\milli\s} to the target input of the DC branch.
Due to its preceding \qty{10}{\Hz} low-pass filter, the edges of the generated pulses are smoothed out, which suppresses high-frequency signal contributions and therefore minimizes the effect of the AC control system on the ramp.
To counteract possible offsets in the DC control branch,~$V_\mathrm{tar}=-10\,\mathrm{mV}$ is chosen as the lower pulse level, ensuring that no light is transmitted before the target signal $V_\mathrm{tar}$ starts to rise.
The resulting intensity signal~$V_\mathrm{out}$ provided by the DC~output of PD-OUT, the low-pass filtered target voltage $\bar{V}_\mathrm{tar}$, and the control signal~$\tilde{V}_\mathrm{c}$ of the AC control system are shown in \autoref{fig: DC Mod}~a).
For the intensity, one finds a $10\,\%$ to $90\,\%$ rise time of \qty{34}{\milli\s} and a fall time of \qty{58}{\milli\s}.
Although the control voltage~$\tilde{V}_\mathrm{c}$ exhibits erratic behavior during the ramps of the target voltage~$V_\mathrm{tar}$, its fluctuations do not significantly affect the light intensity.

To further quantify the impact of the AC control branch on the rising and falling edges of the light intensity, the PD-OUT output signal~$V_\mathrm{out}^{\mathrm{AC}}(t)$ is recorded for various settings of the rise and fall times of the applied ramps, ranging from \qty{0}{\milli\s} to \qty{100}{\milli\s}.
Here, a ramp time of \qty{0}{\milli\s} corresponds to the application of a step function.
For reference, the same measurement is repeated while the AC control system is deactivated, resulting in a different time dependence~$V_\mathrm{out}^{\mathrm{DC}}(t)$ of the power level.
A metric for the influence of the AC control branch on the intensity ramps is the RMSD between the recorded waveforms~$V_\mathrm{out}^{\mathrm{AC}}(t)$ and~$V_\mathrm{out}^{\mathrm{DC}}(t)$.
For its calculation, the two signals are normalized to range between 0 and 1 at the lower and higher setpoint levels, and only data points between the first crossing of the 10\,\% (90\,\%) level and the last crossing of the 90\,\% (10\,\%) level of the rising (falling) edges are considered.
The results of the measurements are shown in \autoref{fig: DC Mod}~b).
For the rising edges, the RMSD is almost independent of the ramp time, since the \qty{10}{\Hz}~low-pass filter causes sufficient smoothing of the setpoint signal~$V_\mathrm{tar}$.
However, for short rise times, the height of the initial spike in the AC control voltage~$\tilde{V}_\mathrm{c}$ (see \autoref{fig: DC Mod}~a)) increases significantly and ultimately reaches \qty{616}{\milli\volt} for a ramp time of \qty{0}{\milli\s} (not shown).
The average RMSD for all rise times is 0.024, indicating the lower measurement limit imposed by noise-induced variations in the signals~$V_\mathrm{out}^{\mathrm{AC}}(t)$ and~$V_\mathrm{out}^{\mathrm{DC}}(t)$.
In contrast, the falling edges are more sensitive to rapid variations in the target signal~$V_\mathrm{tar}$.
For fall times below~\qty{20}{\milli\s}, the RMSD increases significantly, reaching values of around 0.1.
Here, the AC control branch loses stability as the control signal~$\tilde{V}_\mathrm{c}$ falls below \qty{0}{\volt} due to the rapid change in intensity, effectively inverting the sign of the actuator response.
This effect mainly affects the falling edges, where the AC branch significantly reduces its control signal~$\tilde{V}_\mathrm{c}$ to counteract the overall decrease in light intensity.
The inset shows both the signal curves~$V_\mathrm{out}^{\mathrm{AC}}(t)$ and~$V_\mathrm{out}^{\mathrm{DC}}(t)$ for an applied fall time of~\qty{0}{\milli\s}.
The oscillations in the intensity signal~$V_\mathrm{out}^{\mathrm{AC}}(t)$ with active AC control are caused by rapid fluctuations in the control voltage~$\tilde{V}_\mathrm{c}$.
This instability could be compensated for by limiting the control signal~$\tilde{V}_\mathrm{c}$ to positive voltage levels.
Alternatively, if faster intensity ramps are to be performed, the cutoff frequencies of the high-pass filters that are involved in the AC control system can be increased at the expense of a degradation of intensity noise reduction in the low-kHz regime.

\section{Summary and Conclusion}
In summary, we have demonstrated the implementation of a MHz-bandwidth, AOM-based laser intensity stabilization system compatible with high-power applications that combines two feedback branches and one feedforward branch.
Non-linearities in the AOM response are successfully compensated, and RIN levels down to \qty{-155}{\dB\per\Hz} at several hundred kHz have been observed.
Even lower noise levels might be achievable by using PDs with less gain to achieve a lower electronic noise floor and a lower shot noise level.
The laser intensity can be dynamically adjusted within several tens of milliseconds, which is sufficient for many applications in AMO physics.
This makes our intensity control system a powerful tool for applications that require very low RIN levels even at high Fourier frequencies, such as optical lattice experiments with light ultracold atoms.

\section{Declarations}
\paragraph{Availability of data and materials}
The experimental data and evaluation scripts that support the findings of this study are available on Zenodo \cite{partes_aom-based_2026}.
\paragraph{Competing interests}
The authors declare that they have no competing interests.
\paragraph{Funding}
We acknowledge funding from the Horizon Europe program HORIZON-CL4-2022-QUANTUM-02-SGA via the project 101113690 (PASQuanS2.1), the Federal Ministry of Education and Research Germany (BMBF) via the project FermiQP (13N15889), the Deutsche Forschungsgemeinschaft within the research unit FOR5522 (Grant No. 499180199) and the Alfried Krupp von Bohlen and Halbach foundation.
\paragraph{Authors' Contribution}
CP: Design, Implementation, Analysis, and Writing.
JA, EH, KK, FK: Implementation and Writing.
CG: Conceptualization, Supervision, and Writing.
All authors have contributed to the writing of the manuscript.
\paragraph{Acknowledgements}
The authors thank Riccardo Forti of the University of Trieste for his extensive assistance in the programming of the linearization algorithm on the FPGA.
We also acknowledge Jennifer Krauter of the University of Stuttgart for sharing previous work on a fast control system using a FALC controller.
Finally, the first author thanks Magnus Rusch for many valuable discussions on the project.

\section{Methods}

\subsection{AOM step response}
\label{ch: AOM response}
\begin{figure}
	\centering \includegraphics{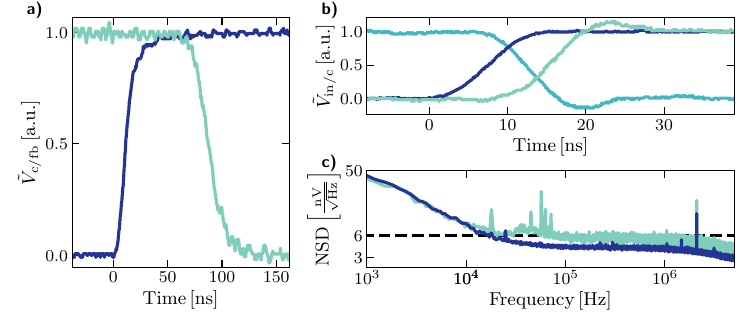}
	\caption{
		a)~Response of the PD-FB2 output voltage~$\tilde{V}_\mathrm{fb}$ (mint green) on a step of the control signal~$\tilde{V}_\mathrm{c}$ (dark blue).
		The data has been scaled to vary between 0 and 1 for better readability. b)~Step response of the feedback (light blue) and the feedforward (mint green) branch of the summing amplifier.
		The corresponding input signal is displayed in dark blue.
		All signals have been offset and normalized to vary between 0 and 1 for better readability. c)~NSD of the summing amplifier output (mint green) on a logarithmic scale.
		For comparison, the noise floor of the measurement device is shown in dark blue. The black dashed line indicates a noise level of \qty{6}{\nano\volt\per\sqrt{\Hz}} as a guide for the eye. }
	\label{fig: aom_sum_amp}
\end{figure}
The path of the laser beam through the AC-AOM was aligned by applying rectangular pulses with an amplitude of~$\tilde{V}_\mathrm{c} = 1\,\mathrm{V}$ to the AOM's control input at the corresponding RF mixer.
The step response of the PD-FB2 signal~$\tilde{V}_\mathrm{fb}$ and the time-averaged transmitted beam intensity were monitored simultaneously, while the two mirrors directly in front of the AOM (see \autoref{fig: Experimental setup}~a)) and the AOM's tilt angle with respect to the beam propagation axis were adjusted.
Additionally, the intensity profile of the laser beam behind the AC-AOM just before the final PBS was recorded with a camera to ensure that no clipping of the laser beam at the edge of the AOM crystal occurred.
The step response of the alignment configuration used throughout this work is presented in \autoref{fig: aom_sum_amp}~a).
For this measurement, the lengths of the signal cables were equalized to avoid offsets caused by propagation delays.
A dead time of less than \qty{65}{\nano\s} with a corresponding 10\,\% to 90\,\% rise time of \qty{35}{\nano\s} was achieved.

\subsection{Summing amplifier}
\label{ch: Summing amplifier}
The summing amplifier used in the described system is a custom-designed signal processing board.
Its main purpose is the analog summation of the fast feedback signal~$\tilde{V}_\mathrm{FALC}$ provided by the FALC, the feedforward signal~$\tilde{V}_\mathrm{ff}$, and the bias offset signal~$V_\mathrm{b}$.
Additionally, the \qty{3}{\kilo\Hz} high-pass RC-filters of the feedback control signal~$\tilde{V}_\mathrm{FALC}$ and the feedforward signal~$\tilde{V}_\mathrm{ff}$ are implemented at the board's corresponding SMA inputs.
Its total output voltage~$\tilde{V}_\mathrm{c}$ is given by~$\tilde{V}_\mathrm{c} = \frac{1}{2}\left(V_\mathrm{b}+K\tilde{V}_\mathrm{ff} - \frac{2}{3}\tilde{V}_\mathrm{FALC}\right)$.
Due to the \qty{50}{\ohm} input termination resistors on the board, the control output is halved with respect to the levels of the incoming signals.
The factor~$K$ is a variable attenuation factor ranging from $0.01$ to $1$, which can be continuously adjusted using a potentiometer to set the control voltage of a voltage-controlled amplifier (Texas Instruments VCA822).
The negative sign in front of the feedback signal~$\tilde{V}_\mathrm{FALC}$ describes its required inversion.
In addition, this signal is attenuated by a factor of~$\frac{2}{3}$ to achieve better noise performance and higher bandwidth for the summing amplifier.
The reaction time of the circuit was designed to be significantly shorter than that of the AC-AOM.
In \autoref{fig: aom_sum_amp}~b), the step responses of the summing amplifier on \qty{100}{\milli\volt}-pulses applied to its feedback and feedforward inputs are shown.
For this measurement, the lengths of the input and output signal cables were equalized to avoid offsets caused by propagation delays.
The dead time of the feedback branch is less than \qty{10}{\nano\s}, corresponding to around 15\,\% of the AOM's reaction time.
In comparison, the feedforward branch is a few ns slower, which can be compensated for by a sufficiently long fiber delay line.
The noise spectral density (NSD) of the summing amplifier output is shown in \autoref{fig: aom_sum_amp}~c).
In the most relevant frequency range above \qty{100}{\kilo\Hz}, the average NSD is around \qty{6}{\nano\volt\per\sqrt{\Hz}}.

\subsection{Low-noise Photodiodes}
\label{ch: Low-noise Photodiodes}
\begin{figure}
	\centering \includegraphics{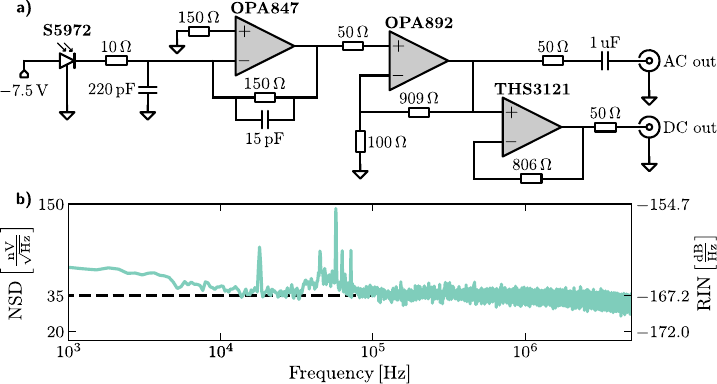}
	\caption{
		a)~Schematic of the signal processing circuit employed for the PDs.
		For simplicity, the design of the corresponding power supply is omitted. b)~Electronic NSD measured at the AC output port of one PD on a logarithmic scale.
		The black dashed line marks the \qty{35}{\nano\volt\per\sqrt{\Hz}}-line as a guide for the eye.
		On the right-hand side, the corresponding RIN for a DC voltage of \qty{8}{\volt} is shown.}
	\label{fig: Photodiode}
\end{figure}
The PDs used in the fast feedback branch (PD-FB2) and the out-of-loop evaluation branch (PD-OUT) are designed to combine a high bandwidth with a low noise floor.
The latter can be approximated by two main noise sources \cite{tricot_power_2018}, namely the electronic background noise~$S_\mathrm{el}(f)$ of the PDs and the shot noise~$S_\mathrm{sn}(f)$ caused by the quantization of electrons into elementary charges.
The shot noise-induced RIN can be calculated by
\begin{equation}
	\label{eq: Shot noise}
	S_\mathrm{sn}(f) = \frac{2e}{I_\mathrm{p}},
\end{equation}
where $e$ is the elementary charge, $f$ the Fourier frequency, and $I_\mathrm{p}$ the induced photocurrent \cite{kwee_new_2011}.
Equation \eqref{eq: Shot noise} implies the necessity of high photocurrents~$I_\mathrm{p}$ and, consequently, high optical powers~$P_\mathrm{opt}$ to reach low RIN levels.

In \autoref{fig: Photodiode}~a), the signal processing circuit employed for the PDs is shown.
The sensor is a high-bandwidth diode (Hamamatsu S5972) with removed glass cover \cite{mazurenko_implementation_2019}, which is biased with a \qty{-7.5}{\volt}-voltage source.
A combination of two operational amplifiers (Texas Instruments OPA847 and OPA892) transforms the photocurrent~$I_\mathrm{p}$ into an output voltage with a total transimpedance gain of around $1.5\,\mathrm{k\Omega}$.
The signal is split and buffered by a current-feedback amplifier (Texas Instruments THS3121) to provide two independent outputs, namely a DC~output port and a faster, high-pass filtered AC~output port (cutoff frequency $\approx3\,\mathrm{kHz}$ for \qty{50}{\ohm} termination).
The maximum output voltage is around \qty{8.5}{\volt}, corresponding to a photocurrent of $I_\mathrm{p}\approx5.6\,\mathrm{mA}$.
To avoid operating the PDs close to their maximum output, the incident laser power~$P_\mathrm{opt}$ was adjusted for typical output voltages of \qtyrange[range-units=single,range-phrase=\,-\,]{7.5}{8.2}{\volt}.
Therefore, the shot noise level is~$S_\mathrm{sn}\approx\,$\qty{-162}{\dB\per\Hz} according to equation \eqref{eq: Shot noise}.

The electronic NSD measured at the AC output port of one PD is presented in \autoref{fig: Photodiode}~b).
It was obtained by recording the PD's output voltage while its photosensitive area was shielded from incident light.
The noise spikes appearing below \qty{100}{\kilo\Hz} are most likely caused by ground loops as similar spikes appear in the NSD of the summing amplifier, see \autoref{fig: aom_sum_amp}~c).
The average noise floor for Fourier frequencies higher than \qty{100}{\kilo\Hz} is \qty{35}{\nano\volt\per\sqrt{\Hz}}, corresponding to a RIN level of \qty{-167}{\dB\per\Hz} for an output voltage of \qty{8}{\volt}.
According to the simple estimate of the out-of-loop lower-RIN limit~$S_\mathrm{min}$ provided in \cite{tricot_power_2018}, $S_\mathrm{min} = 2(S_\mathrm{el}+S_\mathrm{sn})$, the RIN level achievable using the described PDs is approximately $S_\mathrm{min}\approx\,$\qty{-158}{\dB\per\Hz}.
\newpage
\bibliography{bibliography.bib}
\end{document}